\documentstyle[12pt]{article}
\begin{document}
\setlength{\baselineskip}{5mm}

\begin{center}
{\large \bf The fine gradings of $sl(3,C)$ and their symmetries\footnote{In: 
{\it Proceedings of XXIII International Colloquium
 on  Group Theoretical Methods in Physics} 
(A.N. Sissakian, G.S. Pogosyan and L.G. Mardoyan, eds.),
JINR Dubna 2002, Vol. 1, pp. 57--61.}
} \\

\vspace{4mm}

M.~Havl\'{\i}\v{c}ek$^{a,c}$, J.~Patera$^b$, 
\underline{E.~Pelantov\'a}$^{a}$,  J. Tolar$^{a,c}$ \\

\vspace{1mm}

{\small
{$^a$ Faculty of Nuclear Sciences and Physical
Engineering, Czech Technical University in Prague, Czech Republic} \\
{ $^b$ Centre de Recherches
Math\'ematiques, Universit\'e de Montr\'eal, Qu\'ebec, Canada}\\
{$^c$  Doppler Institute of FNSPE, Czech Technical
University, Prague, Czech Republic}
}
\end{center}
\vspace{4mm}

\begin{abstract}
We describe the normalizers for all non-conjugate maximal
 Abelian subgroups of  diagonalizable automorphisms of $sl(3,C)$
 and show their relation to the symmetries of  equations
 related to the graded contraction.
\end{abstract}

%------------------------  Beginning of Main Text ----------\bigskip

 \noindent{\large\bf{Introduction}}

\bigskip

|

Admissible gradings of a simple Lie algebra $L$ over the
 complex or real number field are basic structural
 properties of each $L$.
Examples  of exploitation of coarse gradings like $Z_2$ 
are easy to find in the  physics literature. Here we are
 interested in the opposite extreme:  
the fine gradings of $L$ which only recently
were described [8,9]. Our aim is to point out 
the interesting symmetries of the fine gradings on 
the example of $sl(3,{C})$.

Such symmetries can be used in the study of graded
 contractions of $L$. Indeed, they are the symmetries of 
the system of quadratic equations for the contraction
 parameters. Graded contractions are
a systematic way of forming from $L$ a family
 of equidimensional Lie algebras which are not isomorphic
 to $L$. An insight into such parameter--dependent families
 of Lie algebras provides a group theoretical tool for
 investigating relations between different physical theories
 through their symmetries.

In this contribution we consider finite dimensional $L$ and
announce specific results for $sl(3,{ C})$ only.

The decomposition $\Gamma : \ L= \bigoplus _{i\in I} L_i$ is
called a {\it grading} if, for any pair of indices $i,j \in I$, there
exists an index $k\in I$ such that $0\neq[L_i,L_j]\subseteq L_k$.
There are infinitely many gradings of a given Lie algebra. We do
not need to distinguish those gradings which can be transformed
into each other. More  precisely, if $\Gamma : L=\bigoplus_{i\in
I}L_i$ is a grading and $g$  is an automorphism of $L$, then
${\tilde{\Gamma}} : L= \bigoplus_{i\in  I} g(L_i)$ is also a
grading. Two such gradings are called  {\it equivalent}.

A grading  $\Gamma : L=\bigoplus_{i\in I}L_i$ is a {\it
refinement} of the grading $\overline{\Gamma} : L=\bigoplus_{i\in
J} \overline{L}_j$ if for any $i\in I$ there exists $j \in J$ such
that ${L}_i\subseteq\overline{L}_j$. A grading which cannot be
properly refined is called {\it fine}.

For construction of a grading of $L$, one can use any diagonalizable
automorphism $g$ from $Aut\, L$. It is easy to see that
decomposition of $L$ into eigenspaces of $g$ is a grading.
Similarly, if one takes a set of commuting automorphisms
 from ${A}ut\, L$ and decomposes $L$ according to all
 automorphisms into
their eigenspaces, a grading of $L$ is obtained as well. 
In order
to describe all inequivalent gradings, one has first to
 describe all fine gradings. Then any other grading can be
 found as a coarsening of a fine one (a process inverse to
 refinement of a grading). In [1], the crucial role of 
MAD--groups in the theory of
fine gradings was shown. Here {\it MAD--group} stands  for 
a {\it maximal
Abelian group of diagonalizable automorphisms of} $L$. 
It was proven in
[1] that if $L$ is a simple Lie algebra over $ C$, 
then 
a grading
$\Gamma$ of $L$ is fine if and only if there exists a MAD-group
${\cal G}$ such that $\Gamma$ is the decomposition into
eigenspaces of automorphisms from ${\cal G}$. Moreover, two fine
gradings $\Gamma_1$ and $\Gamma_2$ are equivalent, if and only if the
corresponding MAD-groups ${\cal G}_1$ and ${\cal G}_2$ are
conjugate in $Aut\, L$.

The idea of graded contractions of Lie algebra was introduced in [2].
If $\Gamma: L=\bigoplus_{i\in I}L_i$ is a grading, one can modify
the commutation relation in $L$ by introducing a set of parameters
$\varepsilon_{ij}=\varepsilon_{ji}$, $i,j\in I$ and redefining
$[x,y]_{new}:= \varepsilon_{ij}[x,y]$ for  any $x\in L_i\,\ y\in
L_j$. In order to preserve the Jacobi identities, the parameters
$\varepsilon_{ij}$ must satisfy a system of quadratic equations
[2--7]. The finer the grading is, the bigger system of equations
has to be solved. If $L$ is decomposed into $k$ subspaces, then
the system has, in general, ${k+1}\choose{2}$ variables $\varepsilon_{ij}$.

For Lie algebras of low dimension, or for coarse gradings, the
solution of the system of grading equations is not difficult to
find. However it becomes rather laborious for higher cases
[3,5,6,12]. The symmetries of the set of grading equations could
turn out to be the  tool for finding the solutions more
efficiently.

As we have already said, any fine grading of a simple Lie algebra
over $ C$ is the decomposition $\Gamma$ of $L$ into eigenspaces of
automorphisms belonging to some MAD--group ${\cal G}\subset
Aut\,L$. For any subspace $L_i$ of the fine grading and any
element $g\in\cal G$ we have $g\,L_i=L_i$, i.e. elements of  a
MAD--group preserve each subspace of the grading corresponding to
the MAD--group $\cal G$ ($\cal G$--grading). Here we are interested
in automorphisms which preserve the $\cal G$--grading of $L$ but
not each of its subspaces separately. Clearly  elements of the
normalizer of the MAD--group in ${\cal A}ut\, L$ have this
property. Recall that the normalizer of $\cal G$ is defined as
follows: ${\cal N}({\cal G}) =\{h\in Aut\,L\mid h^{-1}{\cal
G}h\subset\cal G\}$. Let $h \in {\cal N}({\cal G})$ and $L_i$ be a
subspace of the fine grading $\bigoplus_{i\in I}L_i$ corresponding
to the MAD--group ${\cal G}$. Since  $h^{-1}fh\in {\cal G}$ for
arbitrarily chosen $f\in {\cal G}$, we can find $g\in{\cal G}$
such that $h^{-1}fh=g$. Applying this automorphism to $L_i$ and
using that $gL_i = L_i$, we obtain $f(hL_i)=hL_i$, i.e. $hL_i$ is
an eigenspace for any automorphism $f\in{\cal G}$, which means
that $hL_i=L_j$ for some index $j\in I$.

Any $h\in{\cal N}({\cal G})$ generates a permutation $\pi$ on
grading indices $\pi :I \mapsto I$, and thus $h$ provides a
substitution $(\varepsilon_{ij}) \mapsto
(\varepsilon_{\pi(i)\,\pi(j)})$ under which the set of solutions
of quadratic contraction equations is invariant.
\medskip

\bigskip

\noindent{\large\bf{MAD--groups of $sl(3,C)$ and their
normalizers}}

\bigskip

The complete classification of MAD--groups for classical Lie
algebras over $C$ is found in [8].  Let us recall the particular
case of the Lie algebra $sl(3,{ C})$ we are interested in. The
group of automorphisms $Aut\, sl(3,{ C})$ consists of inner
automorphisms $Ad_AX$ and the outer ones $Out_AX$:
 $$
 \begin{array}{rll} 
 Ad_AX &:=A^{-1}XA, \quad & A\in SL(3,{C}) = 
 \{B
\in {C}^{3\times 3}\,, \det B=1\},\\ 
 Out_AX
&:=-(A^{-1}XA)^T\,,\quad &A\in SL(3,{C})
 \end{array}
$$

 According to [8], $Aut\,sl(3,{ C})$ contains four
non-conjugate MAD--groups and therefore four inequivalent fine
gradings. Let us list the four grading groups: 
$$
{\cal
G}_1=\{Ad_A\ |\ A=diag(\alpha_1 ,\alpha _2, \alpha_3), 
\alpha _i
\in
C^*\}~~~~~~~~~~~~~~~~~~~~~~~~~~~~~~~~~~~~~~~~~~~~~~~~~~~~~~~~~~~~$$
$${\cal G}_2=\{Ad_A\ |\ A=diag(\varepsilon _1, \varepsilon _2,
\varepsilon _3),\ \varepsilon _i=\pm 1\}\cup \{ Out _A\ |\
A=diag(\varepsilon _1, \varepsilon _2, \varepsilon _3),\
\varepsilon _i=\pm 1\}~~~~~~~~~~~~~~$$

 $${\cal G}_3=\{Ad_A\ |\
A=diag(\varepsilon , \alpha , \alpha ^{-1}),\ \varepsilon =\pm 1,
\alpha \in C^*\}\cup
~~~~~~~~~~~~~~~~~~~~~~~~~~~~~~~~~~~~~~~~~~~~~~$$
$$\phantom{.}~~~~~~~~~~~~~~~~~~\cup \left\{ Out _A\ |\
A=\left(\begin{array}{ccc}
         \varepsilon &0&0\\
         0&0&\alpha \\
         0&\alpha ^{-1}&0
\end{array}\right),\ \varepsilon = \pm 1, \alpha \in C^*\right\} $$
 $${\cal G}_4 =\left\{Ad_{P^kQ^j}\ |\
Q=\left(\begin{array}{ccc}
        0&1&0\\
        0&0&1\\
        1&0&0
        \end{array}\right),\
P=diag(1, e^{i\frac{2\pi}{3}},e^{i\frac{4\pi}{3}}), \ k,j=0,1,2
\right\}~~~~~~~~~~~~~~~~~~~~~~~~~~~~~~~~$$

Note that ${\cal G}_1$ and  ${\cal G}_4$ are MAD--groups formed by
inner automorphisms only. The group ${\cal G}_1$ is infinite;
it is a maximal torus of $sl(3,{ C})$. The group ${\cal G}_4$ is
finite and consists of 9 elements; it is called the Pauli group in
[10]. The groups ${\cal G}_2$ and ${\cal G}_3$ contain outer
automorphisms.
  There are 8 elements in ${\cal G}_2$. The group
${\cal G}_3$ is infinite, containing a one--parameter subgroup of
the torus. Using these MAD--groups we have decomposed $sl(3,{ C})$
and we have obtained the following fine gradings.

The fine grading corresponding to ${\cal G}_1$ is the Cartan
decomposition consisting of 7 subspaces: one of them is
two-dimensional (the Cartan subalgebra), and 6 remaining subspaces
are one-dimensional. We denote by $E_{ij}\in{ C}^{3\times 3}$ the
matrix with $1$ on the position $ij$ and zeros elsewhere, and by
$\{{E_{ij}}\}_{lin}$ the linear envelope of $E_{ij}$, and we
identify the grading subspaces by the corresponding roots as
subscripts. The Cartan grading has the form:
 $$ \Gamma_1:\
sl(3,{C}) =  N_0\oplus N_{\alpha_1}\oplus N_{\alpha_2} \oplus
N_{\alpha_1+\alpha_2}\oplus N_{-\alpha_1-\alpha_2} \oplus
N_{-\alpha_2}\oplus N_{-\alpha_1}\ , $$
 where
 $N_0=\{diag(a,-a,b,-b)\ |\ a,b\in{C}\}$ and
 $$ \begin{array}{rlrlrl} N_{\alpha_1}&=\{E_{12}\}_{lin}\,,\quad&
N_{-\alpha_1}&=\{E_{21}\}_{lin}\,,\quad&
N_{\alpha_1+\alpha_2}&=\{E_{13}\}_{lin}\,,\\
N_{-\alpha_1-\alpha_2}&=\{E_{31}\}_{lin}\,,\quad&
N_{\alpha_2}&=\{E_{23}\}_{lin}\,,\quad&
N_{-\alpha_2}&=\{E_{32}\}_{lin}\,.
\end{array}$$

 Subspaces in the remaining
 gradings cannot be indexed by roots. Nevertheless,  another nice
choice of the index set
$I$ is possible. Since for each pair of indices $i,j\in I$
 there exists an index $k\in I$ such that $[L_i,L_j] \subseteq
 L_k$, we have a partially binary operation $(i,j)\mapsto k$ on the
index set $I$.
 (If $[L_i,L_j]=\{0\}$, then $k$ can be chosen arbitrarily.)
  It is
proven in [1] that, for simple $L$, the index set $I$ 
  with this operation can be
embedded into
an Abelian additive group $G$. 
  This enables to compute the commutation
 very easily because $[L_i,L_j]\subseteq L_{i+j}$.

For ${\cal G}_1$, $G = Z_{3} \times Z_{3}$ or $Z_7$ [3].

 The index set of the fine grading corresponding to the
group ${\cal G}_2$ can be embedded into the Abelian group
$Z_2\times Z_2\times Z_2$. This group has 8 elements, while the
fine grading has 7 subspaces only: there is no subspace labeled by
the neutral element $(0,0,0)$. The grading has the form:
 $$\Gamma_2: sl(3,C) =  K_{(0,0,1)}\oplus K_{(1,1,1)}\oplus
K_{(1,0,1)}\oplus
 K_{(0,1,1)} \oplus
 K_{(1,1,0)}\oplus  K_{(0,1,0)} \oplus K_{(1,0,0)}\ ,$$
\noindent where $K_{(0,0,1)}=\{diag(a,b,-a-b)\ |\ a,b\in C\}$ and
$$\begin{array}{lll}
 K_{(1,1,1)}=\{E_{21}+E_{12}\}_{lin},\qquad &
K_{(1,0,1)}=\{E_{31}+E_{13}\}_{lin},\qquad &
K_{(0,1,1)}=\{E_{23}+E_{32}\}_{lin},\\
K_{(1,1,0)}=\{E_{21}-E_{12}\}_{lin},\qquad &
K_{(0,1,0)}=\{E_{23}-E_{32}\}_{lin},\qquad &
K_{(1,0,0)}=\{E_{31}-E_{13}\}_{lin} \,.
\end{array}
$$

  The remaining two fine gradings decompose $sl(3,C)$
into 8 one-dimensional subspaces. The grading corresponding to the
MAD--group ${\cal G}_3$ has as its index set  the group $Z_8$:
$$\Gamma_3: sl(3,C) = M_0 \oplus M_1\oplus M_2\oplus M_3\oplus M_4
\oplus M_5\oplus  M_6 \oplus M_7\, ,$$
 where
$$\begin{array}{lll} M_0=\{E_{22}-E_{33}\}_{lin},\quad &
M_1=\{E_{12}-E_{31}\}_{lin},\quad &
 M_2=\{E_{23}\}_{lin},\\
 M_3=\{E_{13}+E_{21}\}_{lin},&
 M_4=\{2E_{11}-E_{22}-E_{33}\}_{lin},\quad&
 M_5=\{E_{12}+E_{31}\}_{lin},\\
 M_6=\{E_{32}\}_{lin},&
 M_7=\{E_{13}-E_{21}\}_{lin}\,. &\\
\end{array}
$$

 The last fine grading corresponding to the
MAD--group ${\cal G}_4$ can be briefly written  in terms of 
generalized Pauli matrices $P$ and $Q$ introduced in the
description of the MAD--group itself. Definition of generalized
Pauli matrices in $C^{n\times n}$ and their properties are found
in [10], [11]. Now the index set is a subset of the group $Z_3\times
Z_3$. Again, no subspace is labeled by the neutral element:
$$\Gamma_4: sl(3,C) =  L_{(1,0)}\oplus L_{(2,0)}\oplus
L_{(0,1)}\oplus L_{(0,2)} \oplus L_{(1,1)}\oplus L_{(2,1)} \oplus
L_{(1,2)}\oplus L_{(2,2)}\ ,$$ 
where
 $$\begin{array}{llll}
L_{(1,0)}=\{P\}_{lin},\quad & L_{(2,0)}=\{P^2\}_{lin},\quad &
L_{(0,1)}=\{Q\}_{lin},\quad & L_{(0,2)}=\{Q^2\}_{lin},\\
L_{(1,1)}=\{PQ\}_{lin},\quad & L_{(2,1)}=\{P^2Q\}_{lin},\quad &
L_{(1,2)}=\{PQ^2\}_{lin},\quad & L_{(2,2)}=\{P^2Q^2\}_{lin}\, .
\end{array}
 $$

Since any grading can be derived from some of the  fine ones by
collecting several subspaces together into  subspaces of higher
dimension, one can construct all coarser gradings, i.e.  try all
possibilities of coarser decompositions and check for each
decomposition  whether it is a grading or not. Let us note that
the algebra $sl(3,{ C})$ has 17 inequivalent gradings including 4
fine gradings and the coarsest one formed by only one space,
namely the whole $sl(3,{ C})$.

For any group ${\cal G}$, the normalizer ${\cal N}({\cal G})$
contains ${\cal G}$ as its normal subgroup. Hence one has the
quotient group ${\cal N}({\cal G})/{\cal G}$. Now two
elements $h_1$ and $h_2$ of the normalizer  ${\cal N}({\cal G})$
belong to the same coset, if there exists $g\in {\cal G}$ such
that $h_1=h_2g$. For the grading $\Gamma$ it means that the
automorphisms $h_1$ and $h_2$ yield  the same permutation of
the
subspaces of $\Gamma$. Therefore ${\cal N}({\cal G})/ {\cal G}$ is
isomorphic to some subgroup of the symmetric group $S_n$, where
$n$ is the cardinality of the index set $I$. An automorphism which
preserves the grading maps a subspace $L_i$ to a subspace
$L_j=gL_i$ of the same dimension. Particularly, for gradings
$\Gamma_1$ and $\Gamma_2$, the two dimensional subspaces $N_0$ and
$K_{(0,0,1)}$  must be mapped onto themselves. Thus
only permutations of the remaining subspaces are possible. There
are $6! = 720$ of such permutations. The normalizers are
in fact
much smaller. The gradings $\Gamma_3$ and $\Gamma_4$ consist of 
one--dimensional subspaces only, hence there are $8!$ permutations of
these subspaces. But again, only very few of
this enormous number of permutations correspond to an
automorphism, as will be seen below.

Let $B_1$ and $B_2$ denote the permutation matrices
 $$
   B_1=\left(\begin{array}{ccc}
             0&1&0\\
             0&0&1\\
             1&0&0
         \end{array}\right) \qquad
         {\rm and }\qquad
    B_2=\left(\begin{array}{ccc}
             1&0&0\\
             0&0&1\\
             0&1&0
         \end{array}\right)
$$ and $H$  the diagonal matrix $H=diag(1,i,i)$.

\medskip

\noindent $\bullet$ {\it The normalizer of the MAD--group} ${\cal
G}_1$ contains besides ${\cal G}_1$ the transposition $Out_I$ and
the automorphisms $Ad_B$, where $B_i$ are permutation matrices.
Inner automorphisms of ${\cal N}({\cal G}_1)/{\cal G}_1$  form the
symmetry group $S_3$. It is easy to check that such automorphisms
belong to the normalizer. To prove that any other element of the
normalizer is only their composition, is only somewhat more
laborious. The conclusion is:
 $$
{\cal N}({\cal G}_1)/{\cal G}_1 \ \ {\rm has\ 12\ elements\ and\
 is\ generated \ by}\ \ Out_I, Ad_{B_1}, Ad_{B_2}\,.
$$

\medskip

\noindent $\bullet$ {\it The normalizer of the MAD--group} ${\cal
G}_2$ contains besides ${\cal G}_2$ the automorphisms  $Ad_{B_1}$,
$Ad_{B_2}$ and $Ad_H$. We thus conclude: 
$$ {\cal N}({\cal
G}_2)/{\cal G}_2 \  {\rm  has\ 18\ elements\ and\
 is\ generated \ by}\   Ad_{B_1}, Ad_{B_2}, Ad_H
.
$$

\medskip

\noindent $\bullet$ {\it The normalizer of the MAD--group} ${\cal
G}_3$ contains besides ${\cal G}_3$ the automorphisms  $Ad_{B_2}$
and $Ad_H$. These commuting automorphisms are of  order two (in the
quotient  group) and generate the normalizer. The conclusion is:
 $$
{\cal N}({\cal G}_3)/{\cal G}_3 \ \ {\rm is\ isomorphic\ to}\
 Z_2\times Z_2,\ {\rm  has\ 4\ elements\ and\ is\ generated \ by}\
 Ad_{B_2}, Ad_H
.
$$

\medskip

\noindent $\bullet$ {\it The normalizer of the MAD--group} ${\cal
G}_4$ needs for its description the Sylvester matrix $S$ and the
diagonal matrix $D=diag(1,1,e^{\frac{2\pi i}{3}})$. The Sylvester
matrix is the symmetric matrix which transforms the generalized
Pauli matrices $P$ and $Q$ into each other, i.e.  $SPS^{-1}=Q$. It
is curious to note that the subgroup of ${\cal N}({\cal G}_4)$
formed by the inner automorphisms is isomorphic to the group
$SL(2,Z_3)$ of order 24. Finally we can conclude:
  $$
{\cal N}({\cal G}_4)/{\cal G}_4 \  {\rm has\ 48\ elements\ and\
 is\ generated \ by}\   Out_I, Ad_S,  Ad_D
$$ 
Unlike the cases of the groups ${\cal G}_1$, ${\cal G}_2$ and
${\cal G}_3$, it is a nontrivial task to show that  the normalizer
of ${\cal G}_4$ is generated by the three automorphisms given above. 
The
detailed proof of this statement can be found in [11] 
and concerns
MAD--groups generated by generalized Pauli matrices 
in the general case of $sl(n,C)$.

\subsection*{Acknowledgements}
M.H., E.P. and J.T. gratefully acknowledge the support of the
Ministry of Education of Czech Republic under the contract MSM
210000018. J.P. has been partially supported by the National
Science and Engineering Research Council of Canada.

%------------------------------- End of Main Text --------------------

\newcommand{\etal}{{\em et al.}}
\setlength{\parindent}{0mm} \vspace{5mm} {\bf References}
\begin{list}{}{\setlength{\topsep}{0mm}\setlength{\itemsep}{0mm}%
\setlength{\parsep}{0mm}}

\item[1.] J.~Patera, H.~ Zassenhaus, Linear Algebra \& Its Appl.
 {\bf 112} (1989) 87--159.
\item[2.] M.~de~Montigny, J.~Patera, J. Phys. A: Math. Gen. {\bf 24} (1991), 525--549.
\item[3.] M.~Couture, J.~Patera, R.~T.~Sharp, P.~Winternitz, J.Math.
Phys.
{\bf 32} (1991) 2310--2318.
\item[4.] M.~de~Montigny, J.~Patera, J.~Tolar,  J. Math. Phys. {\bf
35} (1994), 405--425.
\item[5.]
  J.~Tolar, P.~Tr\'avn\'{\i}\v{c}ek, J. Math. Phys. {\bf 36} (1995),
4489--4506.
\item[6.] J.~Tolar, P.~Tr\'avn\'{\i}\v{c}ek, J. Math. Phys. {\bf 38}
(1997), 501--523.
\item[7.] J.~Tolar, in "Lie Theory and Its Applications in Physics",
 (H.-D. Doebner, V.K. Dobrev and J. Hilgert, editors)
 World Scientific, Singapore 1996, pp. 226--238.
\item[8.]  M.~Havl\'{\i}\v{c}ek, J.~ Patera, E.~Pelantov\'a,
 Linear Algebra \& Its Appl. {\bf 277} (1998) 97--125.
\item[9.]  M.~Havl\'{\i}\v{c}ek, J.~Patera, E.~Pelantov\'a,
 Linear Algebra \& Its Appl. {\bf 314} (2000) 1--47.
 \item[10.] J.~Patera, H.~Zassenhaus, J. Math. Phys. {\bf 29} (1988)
 665--673.
 \item[11.] M.~Havl\'{\i}\v{c}ek, J.~Patera, E.~Pelantov\'a,
 J.~Tolar, Automorphisms of the fine grading of $sl(n,C)$ 
   associated with the generalized Pauli
 matrices,
   J. Math. Phys. {\bf 43} (2002), 1083--1094.
 \item[12.] M.~A.~Abdelmalek, X.~Leng, J.~Patera, P.~Winternitz,
 J. Phys. A: Math. Gen. {\bf 29} (1996) 7519--7543.

\end{list}
\end{document}